\documentclass[preprint,12pt]{elsarticle}

\usepackage{amssymb}

\journal{a journal}

\begin{document}

\begin{frontmatter}



\title{Limitation of network inhomogeneity in improving cooperation in coevolutionary dynamics}


\author {Li-Xin Zhong$^a$}\ead{zlxxwj@163.com}
\author {Tian Qiu$^b$}

\address[label2]{School of Journalism, Hangzhou Dianzi University, Hangzhou, 310018, China}
\address[label3]{School of Information Engineering, Nanchang Hangkong University, Nanchang, 330063, China}

\begin{abstract}
Cooperative behavior is common in nature even if selfishness is sometimes better for an individual. Empirical and theoretical studies have shown that the invasion and expansion of cooperators are related to an inhomogeneous connectivity distribution. Here we study the evolution of cooperation on an adaptive network, in which an individual is able to avoid being exploited by rewiring its link(s). Our results indicate that the broadening of connectivity distribution is not always beneficial for cooperation. Compared with the Poisson-like degree distribution, the exponential-like degree distribution is detrimental to the occurrence of a higher level of cooperation in the continuous snowdrift game (CSG).
\end{abstract}


\end{frontmatter}


\section{Introduction}\label{sec:introduction}
Understanding the existence of cooperation in benefit-seeking biological systems and human societies is one of the most fascinating problems studied by biologists, physicists and sociologists\cite{nowak,wang,zhong1,zheng}. Traditionally, penalty-based approach had been taken as the dominant mechanism for the maintenance of cooperation\cite{fu,hauert}. Until recently, to mimic the limited partnership in individual interactions, population structures are introduced into the evolutionary process. Among them, the regular network, the small world network and the scalefree network are the most popular static networks\cite{watts,albert,ivanov,masucci}. On a static network, mutual interactions can only occur between the individuals with immediate and fixed connections. Depended upon the prisoner's dilemma (PD) and the snowdrift game (SG)\cite{baek,ohtsuki,szolnoki,zhong2}, the role of topological properties in the invasion and expansion of cooperation have been discussed a lot\cite{chan,masuda,tanimoto,perc1,perc2}.

In real society, both the level of cooperation and the individual connection are continuously evolving\cite{johnson,xu,zhong3}, such as that in the ultimatum game\cite{eguiluz,sinatra,li,iranzo}, two interacting agents, called proposer and responder respectively, share a given amount of money with a continuously changing ratio, and other social dilemma games with a continuously changing linkage\cite{szolnoki1,szolnoki2,szolnoki3,perc3}. In the present work, to address the effects of dynamic linkage on the change of cooperative behavior, a kind of cost-benefit game, called the continuous snowdrift game (CSG)\cite{doebeli,zhong4}, and an adaptive network model can be combined to serve as a practical tool.

In the CSG, a continuous variable x represents the level of cooperation, such as the amount of money invested in a company or the time spent in public affairs. The investment x not only benefits to the recipient but also the donor itself. For a pairwise interaction between agent i who invests x and agent j who invests y, the payoffs to i and j can be expressed as $P(x,y)=B(x+y)-C(x)$ and $P(y,x)=B(y+x)-C(y)$ respectively, in which $B(x)$ and $C(x)$ represent the benefit and cost functions. Mutation and selection occasionally occur in the evolutionary process. The evolution of x is governed by the invasion fitness $f(y)=P(y,x)-P(x,x)$, which represents the invasion ability of a rare mutant y into a population with a monomorphic investment x.

On the adaptive network, personal preference and a potential partner's reputation are both very important for an agent to make a relinking decision\cite{zhong5,zschaler,chen}. For example, as an agent is ready to do business, he may choose a cooperative investor who is willing to invest more or an experienced investor who has made more profits from his former investment as his partner. In this paper, we incorporate two kinds of agents, the agents with investment-dependent preference (IDP) and the agents with wealth-dependent preference (WDP), into the original CSG. In the rewiring process, an IDP agent tends to reconnect to a generous investor and a WDP agent tends to reconnect to the rich. We have two main findings in the present work.

(1) Compared with that on the static network, in the present model, cooperative behavior is highly promoted on the dynamic network. The rewiring process promotes the formation of highly-investing clusters which helps the cooperators defeat the defectors.

(2) In comparison with the Poisson-like degree distribution, the exponential-like degree distribution is detrimental to the occurrence of a higher level of cooperation. A theoretical analysis indicates that, on the network with an exponential-like degree distribution, it is the inhomogeneous connectivity that leads to the unstableness of cooperator clusters and thereafter a decrease in average investment.

The paper is organized as follows. In Section \ref{sec:model}, the CSG model with IDP and WDP agents and related variables are introduced. In Section \ref{sec:results}, simulation results are presented and discussed in detail. Analytical calculations are given in Section \ref{sec:analysis} and conclusions are summarized in the last section.

\section{The model} \label{sec:model}

In the present model, there are N agents who are initially arranged on the nodes of a random regular network (RRG) with average degree $<k>$. Each agent has a pre-defined investment x which is a real variable between 0 and 1 and a specific preference for switching, IDP or WDP. The agents with immediate connections make pairwise interactions and each agent's investment and linkage evolve according to the following procedures.

Step1: A randomly chosen agent i interacts with all its immediate neighbors and a summed payoff is attained $P_{i}=\sum_{j=1}^{k_{i}}[B(x_{i}+x_{j})-C(x_{i})]$, where $k_{i}$ is the degree of agent i and the benefit and cost functions satisfy $B(x)=b_{2}x^2+b_{1}x$ and $C(x)=c_{2}x^2+c_{1}x$ respectively.

Step 2: Agent i compares its payoff with one of its immediate neighbors agent j's and gets its fitness $f_{i}=P_{i}-P_{j}$. If $f_{i}<0$, with probability $\lambda=-\frac{f_{i}}{\alpha}$, in which $\alpha=max(k_{i},k_{j})(B_{max}-C_{min})$ is a normalization factor used to ensure $0\leq \lambda \leq1$, agent i adopts the investment the same as agent j's. Or else, $\lambda=0$.

Step 3: Occasionally, agent i replaces its investment with a mutant, which is drawn randomly from a Gaussian distribution with its mean $x_{i}$ and variance $\sigma^2$. Just as that in ref.\cite{doebeli}, in this paper we have the mutation probability $\mu=0.01$ and the standard deviation $\sigma=0.005$.

Step 4: The payoffs for agent i in the latest T time steps are accumulated as wealth, that is, $w_{i}=\sum_{t'=t-T+1}^{t}P_{i}(t')$. If $w_{i}$ is less than the average wealth $<w>=\frac{\sum_{i=1}^{N}w_{i}}{N}$, agent i rewires its link(s) with probability $\psi\in[0,1]$. In the reconnection process, an agent chooses its new partner according to its attachment preference, IDP or WDP. The ratio of IDP (or WDP) agents in the population is $P_{IDP}$ (or $P_{WDP}$) , $P_{IDP}+P_{WDP}=1$. An IDP agent chooses its new partner according to its investment, that is, an agent j may be chosen as agent i's new partner with probability $x_{j}$. A WDP agent chooses its new partner according to its wealth, that is, an agent j may be chosen as agent i's new partner with probability $\frac{w_{j}+\varepsilon}{\sum_{i=1}^{N}(w_{i}+\varepsilon)}$, where $\varepsilon=0.01$ is used to avoid $\sum_{i=1}^{N}(w_{i}+\varepsilon)=0$. If an agent's degree becomes 0 because of rewiring, it will randomly choose an agent as its partner while the partner will randomly cut one of its links. For each agent, the judgement about rewiring or not is made every T time steps.

In the original CSG\cite{doebeli}, the evolved investment x is governed by the selection gradient $G(x)=\frac{\partial f(y)}{\partial y}\mid_{y=x}$. The singular investment x* is the solution of equation $G(x)=0$, that is, $B'(2x)-C'(x)=0$. With quadratic benefit and cost functions, the singular investment becomes $x^*=\frac{c_{1}-b_{1}}{4b_{2}-4c_{2}}$. If $\frac{dG(x)}{dx}\mid_{x=x^*}<0$, the singular gene value x* is stable and evolutionary branching occurs for $2b_{2}<c_{2}<b_{2}<0$. If there is no solution for $B'(2x)-C'(x)=0$ within the domain, x increases or decreases monotonically to the boundary.

In the present work, the network with degree distribution $P(k)=\delta(k-<k>)$ is referred to as a homogeneous network and the network with a broader degree distribution an inhomogeneous network. We are mainly concerned about the role of network structure in the change of evolutionary branching and the improvement of cooperation. Therefore, throughout the paper, we adopt the benefit and cost functions used in ref.\cite{doebeli}, that is, $B(x)=-1.4x^2+6x$, $C(x)=-1.6x^2+4.56x$, with which evolutionary branching occurs, and $B(x)=-1.5x^2+7x$, $C(x)=-x^2+8x$, with which the average investment decreases to zero in the well-mixed case.

\section{Results and discussions}\label{sec:results}
We start by investigating how the rewiring process affects the evolutionary branching and the average investment. Throughout the paper, the total number of agents is N=5000 and the average degree of nodes is $<k>=6$.

\begin{figure}
\includegraphics[width=10cm]{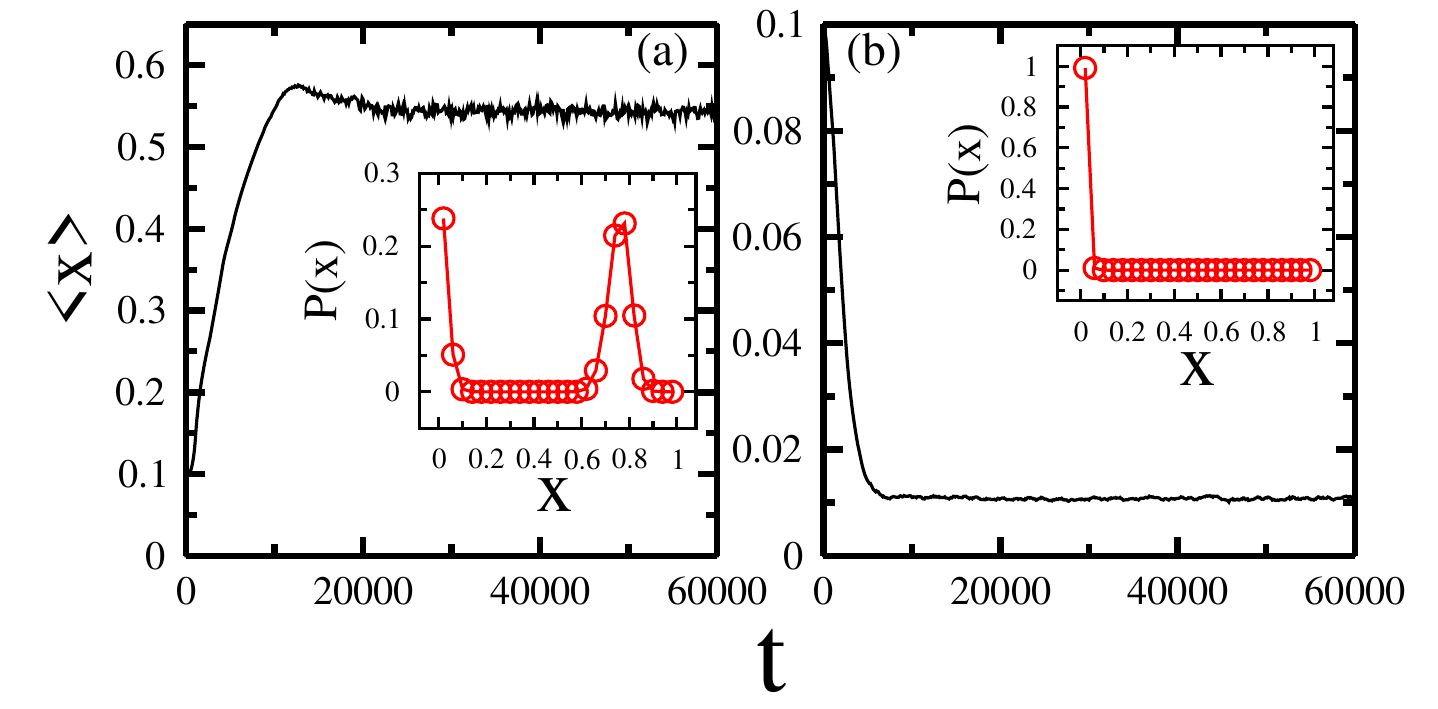}
\caption{\label{fig:epsart}Average investment versus time on a RRG network. The inset shows the corresponding investment distribution. (a)$B(x)=-1.4x^2+6x$, $C(x)=-1.6x^2+4.56x$; (b)$B(x)=-1.5x^2+7x$, $C(x)=-x^2+8x$.}
\end{figure}

Figure 1(a) and (b) show the temporal average investment $<x>$ in the RRG. Compared with the results in the well-mixed case\cite{doebeli}, in the static network with degree distribution $P(k)=\delta(k-<k>)$, the structured process has little effect on the system behavior. For $b_{2}=-1.4$, $b_{1}=6$, $c_{2}=-1.6$, $c_{1}=4.56$, which satisfy $2b_{2}<c_{2}<b_{2}<0$, evolutionary branching occurs. The two clusters evolve to $x\sim0.01$ and $x\sim0.8$ and the average investment reaches $<x>\sim0.55$. For $b_{2}=-1.5$, $b_{1}=7$, $c_{2}=-1$, $c_{1}=8$, the average investment decreases monotonically to $x\sim0.01$.

\begin{figure}
\includegraphics[width=10cm]{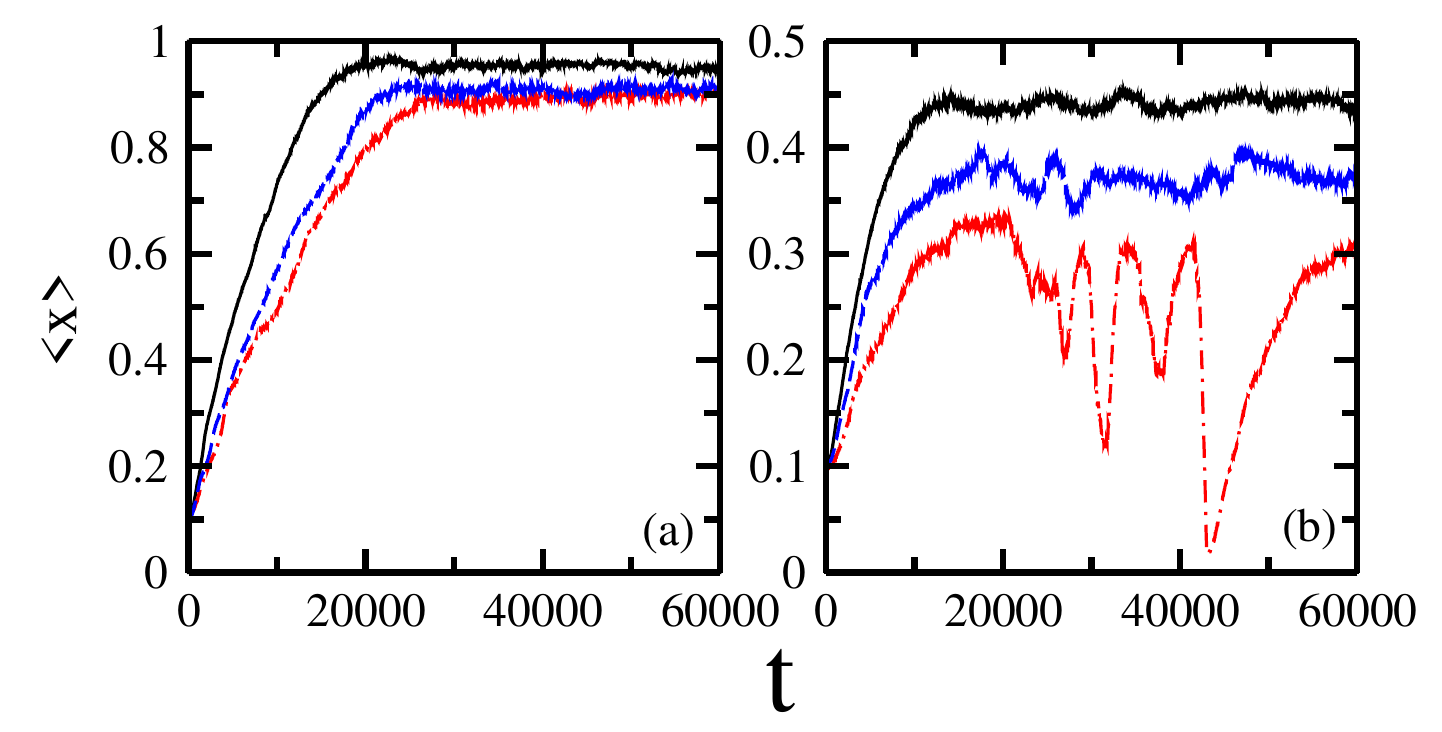}
\caption{\label{fig:epsart}Average investment versus time on a dynamic network for $\psi=0.5$ and $P_{IDP}=1$ (black), 0.5 (blue), 0 (red). (a)$B(x)=-1.4x^2+6x$, $C(x)=-1.6x^2+4.56x$; (b)$B(x)=-1.5x^2+7x$, $C(x)=-x^2+8x$.}
\end{figure}

As we incorporate the rewiring mechanism into the above evolutionary process, the system behavior changes a lot. In Fig.2 (a) and (b) we give the temporal average investment in the dynamic network for $\psi=0.5$ and different $P_{IDP}$ (or $P_{WDP}$). Other values of $\psi$ have also been tried in the simulation and it is observed that the change of $\psi$ only affects the relaxation time and the average investment but not the advantage of some kind of relinking preference. Figure 2 (a) and (b) show that the rewiring process can highly promote cooperation in the coevolutionary CSG. In comparison with the results in Fig.1(a) and (b), the average investment increases obviously as the two kinds of reconnection preferences are incorporated into the rewiring process. As we consider the role of the ratio of IDP (or WDP) agents in the population, we find that the existence of more IDP agents can make the system reach a higher level of cooperation than that with more WDP agents. In addition to that, from Fig.2(b) we also find that more WDP agents will result in a large fluctuation in the average investment.

\begin{figure}
\includegraphics[width=10cm]{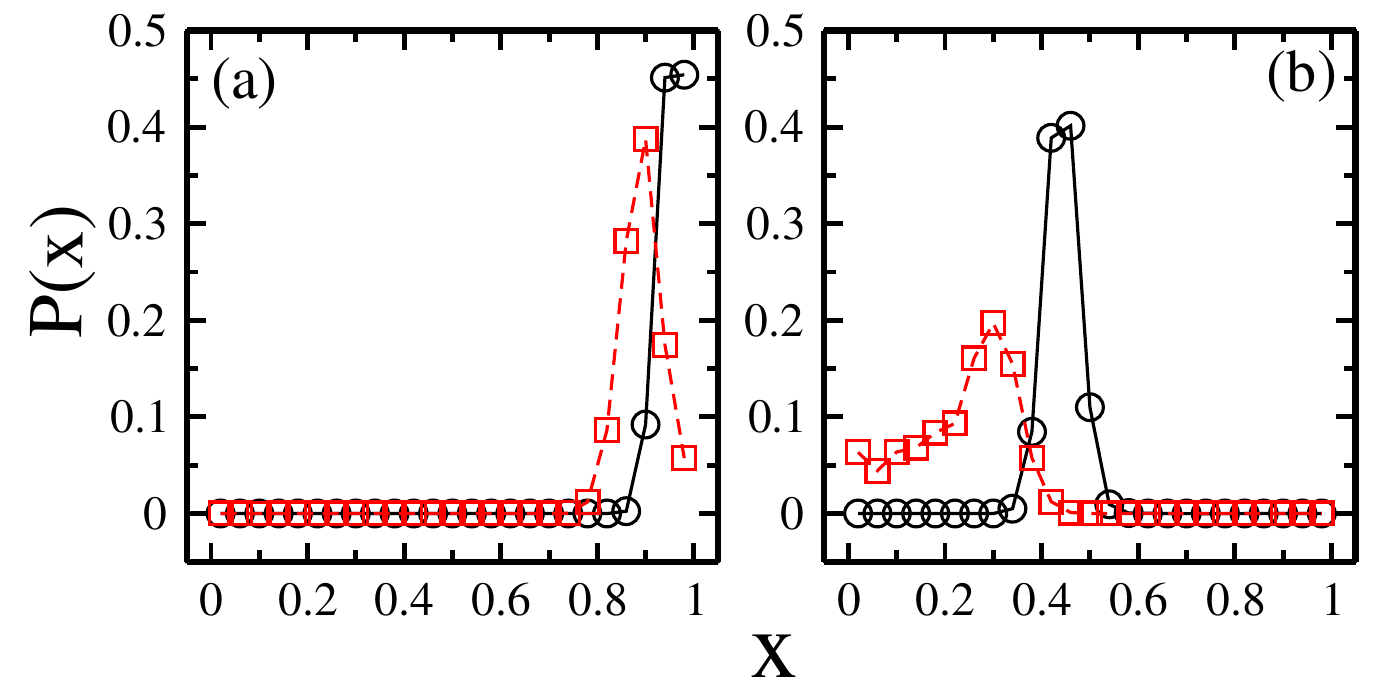}
\caption{\label{fig:epsart}The investment distribution in the final steady state for $\psi=0.5$ and $P_{IDP}=1$ (circles), 0 (squares). Averaged over 10 runs and in each run the data are obtained by averaging over 10000 time steps after 50000 relaxation time.  (a)$B(x)=-1.4x^2+6x$, $C(x)=-1.6x^2+4.56x$; (b)$B(x)=-1.5x^2+7x$, $C(x)=-x^2+8x$.}
\end{figure}

To have a close eye of the role of rewiring in the change of individual investment, in Fig.3(a) and (b) we give the investment distribution after 50000 relaxation time. Figure 3 (a) shows that the rewiring process leads to the disappearance of evolutionary branching observed in the RRG.  In the final steady state, nearly all the agents make the same investment on the dynamic network. From Fig.3(b) we observe that, for a large $P_{IDP}$, the investment distribution is like a delta function. But for a small $P_{IDP}$, it becomes broader. From the time dependent distribution we find that such a broadened distribution results from the large fluctuation of the investment.

\begin{figure}
\includegraphics[width=10cm]{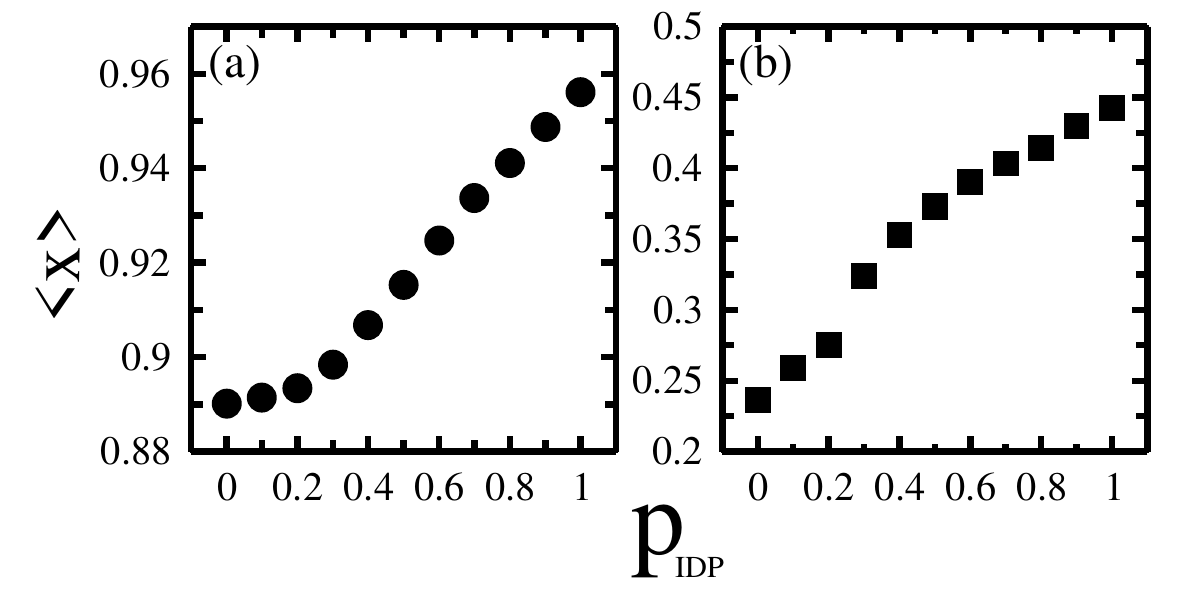}
\caption{\label{fig:epsart}The averaged investment in the final steady state as a function of $P_{IDP}$ for $\psi=0.5$. Averaged over 10 runs and in each run the data are obtained by averaging over 10000 time steps after 50000 relaxation time.  (a)$B(x)=-1.4x^2+6x$, $C(x)=-1.6x^2+4.56x$; (b)$B(x)=-1.5x^2+7x$, $C(x)=-x^2+8x$.}
\end{figure}

Figure 4 (a) and (b) show the averaged investment in the final steady state as a function of $P_{IDP}$. It is observed that $<x>$ increases monotonically with the rise of $P_{IDP}$ in both cases. As $P_{IDP}$ increases from 0 to unity, $<x>$ increases from 0.89 to 0.95 in Fig.4(a) while $<x>$ increases from 0.24 to 0.44 in Fig.4(b). For $N\gg1$, the change of population size has little effect on the change of $<x>$.

\begin{figure}
\includegraphics[width=5cm]{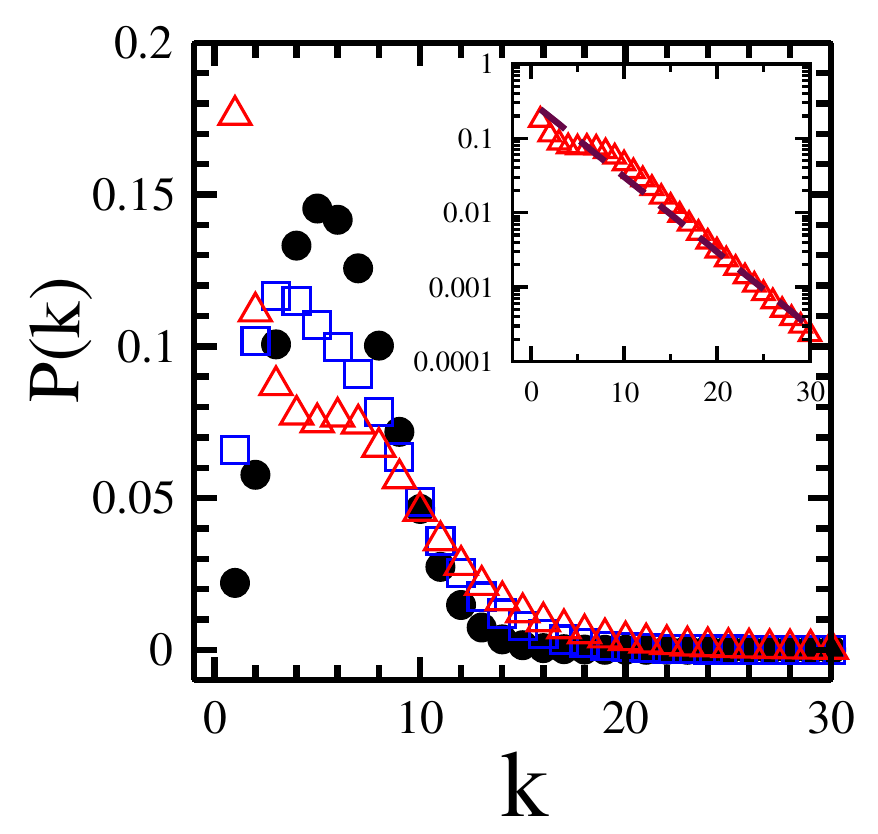}
\caption{\label{fig:epsart}The degree distribution of the evolved network for $\psi=0.5$ and $P_{IDP}=1$ (circles), 0.5 (squares), 0 (triangles). Averaged over 10 runs and in each run the data are obtained by averaging over 10000 time steps after 50000 relaxation time. $B(x)=-1.4x^2+6x$, $C(x)=-1.6x^2+4.56x$. Inset: P(k) for $P_{IDP}=0$ (triangles) and the fitting curve $P(k)=0.311e^{-0.232k}$}
\end{figure}

Because of the unequal wealth distribution, the inter-cluster connections are continuously removed and reconnected as the investment evolves. To find the effect of adaptive rewiring on the emergence of a typical network structure, in Fig.5 we plot the degree distribution of the evolved network in the final steady state. All the parameters are the same as that in Fig.2(a). In the system where all the individuals tend to set up new connections with the agents with a larger investment (corresponding to $P_{IDP}=1$), P(k) evolves to a Poisson-like degree distribution. But in the system where all the individuals tend to set up new connections with the agents with larger wealth (corresponding to $P_{IDP}=0$), P(k) evolves to an exponential-like degree distribution. As we compare the results in Fig.2(a) with that in Fig.5, we find that, compared with an exponential-like degree distribution, a Poisson-like degree distribution is more beneficial for the occurrence of a higher level of cooperation in the present model.

\begin{figure}
\includegraphics[width=10cm]{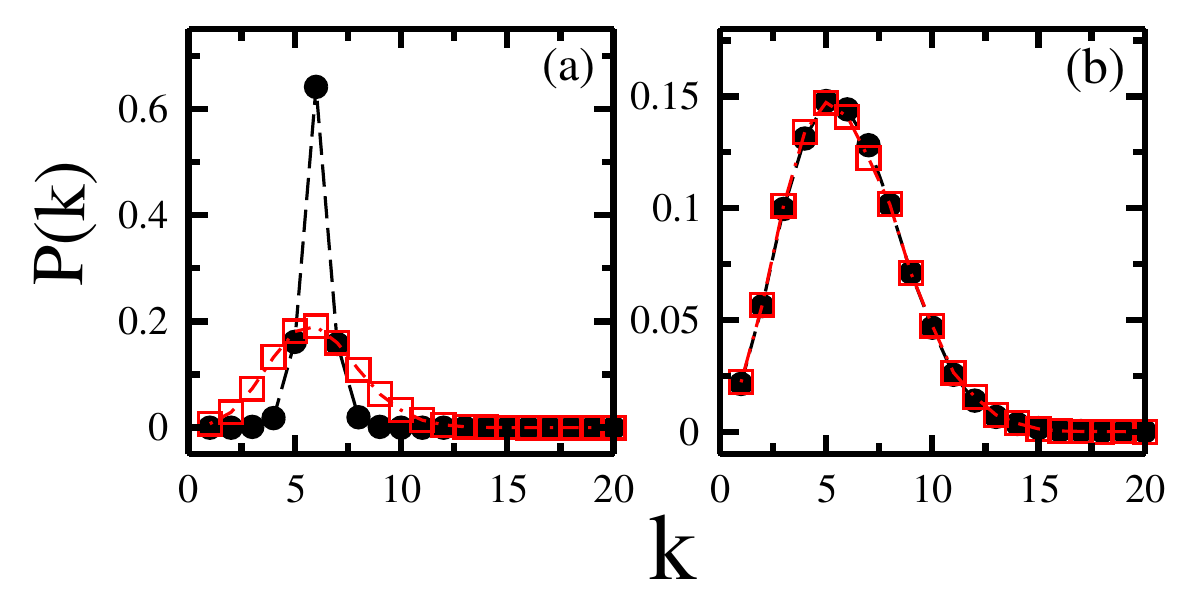}
\caption{\label{fig:epsart}The degree distribution in the FDTS network for $P_{IDP}=1$ and $B(x)=-1.4x^2+6x$, $C(x)=-1.6x^2+4.56x$. (a) $\psi=0.001$, $\tau=1000$(circles), 30000(squares); (b)$\psi=0.5$, $\tau=1000$(circles), 30000(squares). Averaged over 10 runs.}
\end{figure}

The above results are quite different from the results found in the static network where an exponential degree distribution is more beneficial for cooperation \cite{assenza}. To understand how such a difference comes from, in the following we set up a firstly-dynamic-then-static (FDTS) network and have a close eye of the relationship between $<x>$ and P(k). The FDTS network is attained as follows: start from a random regular network, at the first $\tau$ time steps, the connection coevolves with the investment just as that in the dynamic network. Then, the network becomes static while the investment evolves further. Figure 6 (a) and (b) show the degree distribution in the FDTS network for different $\psi$ and $\tau$. For a small rewiring probability $\psi=0.001$, P(k) becomes broader as $\tau$ increases from 1000 to 30000. But for $\psi=0.5$, the change of $\tau$ has little effect on the change of P(k). A broader P(k) means a larger difference in degree between the agents. Therefore, Figure 6 indicates that the inhomogeneity of individual connections increases with the rise of $\tau$ for a small $\psi$ but not for a large $\psi$. Other values of $\psi>0.01$ are also tried in the present model, compared with that in Fig.6(b), it is found that the change of $\psi$ only change the relaxation time but not the degree distribution in the final steady state.

\begin{figure}
\includegraphics[width=10cm]{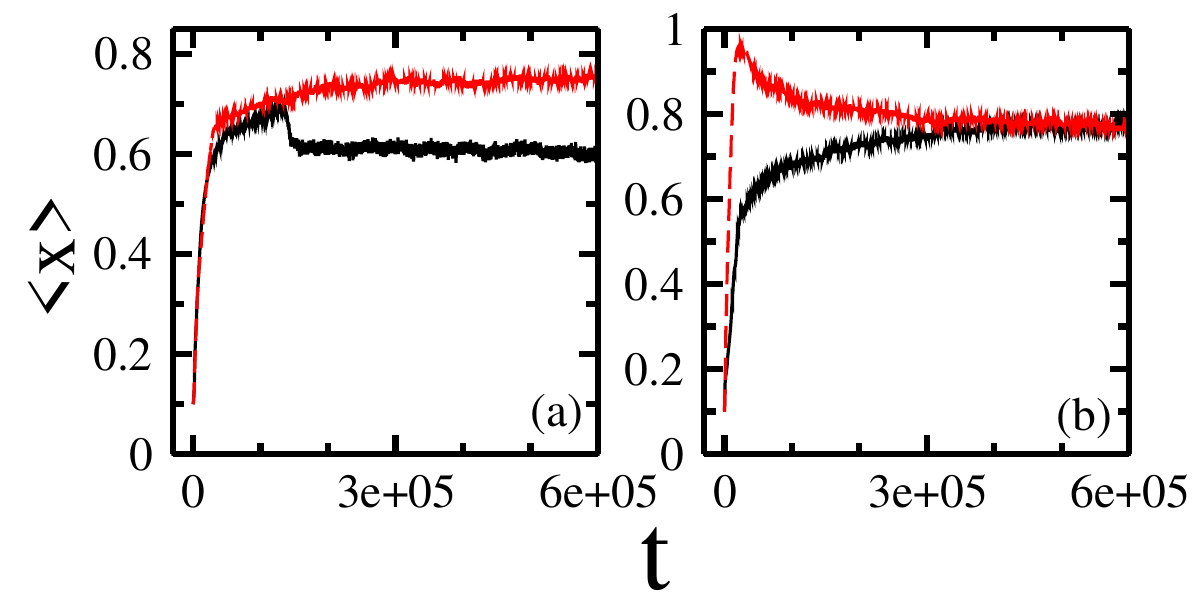}
\caption{\label{fig:epsart}Average investment versus time in the FDTS network for $P_{IDP}=1$ and $B(x)=-1.4x^2+6x$, $C(x)=-1.6x^2+4.56x$. (a) $\psi=0.001$, $\tau=1000$(black), 30000(red); (b)$\psi=0.5$, $\tau=1000$(black), 30000(red).}
\end{figure}

Accordingly, in Fig.7 (a) and (b) we plot the temporal average investment in the FDTS network. With rewiring probability  $\psi=0.001$, as $\tau$ increases from 1000 to 30000, the average investment increases from 0.6 to 0.75. However, with rewiring probability $\psi=0.5$, the change of $\tau$ does not change the average investment in the final steady state but the relaxation time. The above results indicate that the system behavior is closely related to the degree distribution in the FDTS network. Just as that in the static network, a broader degree distribution is beneficial for cooperation in the FDTS network.

\begin{figure}
\includegraphics[width=10cm]{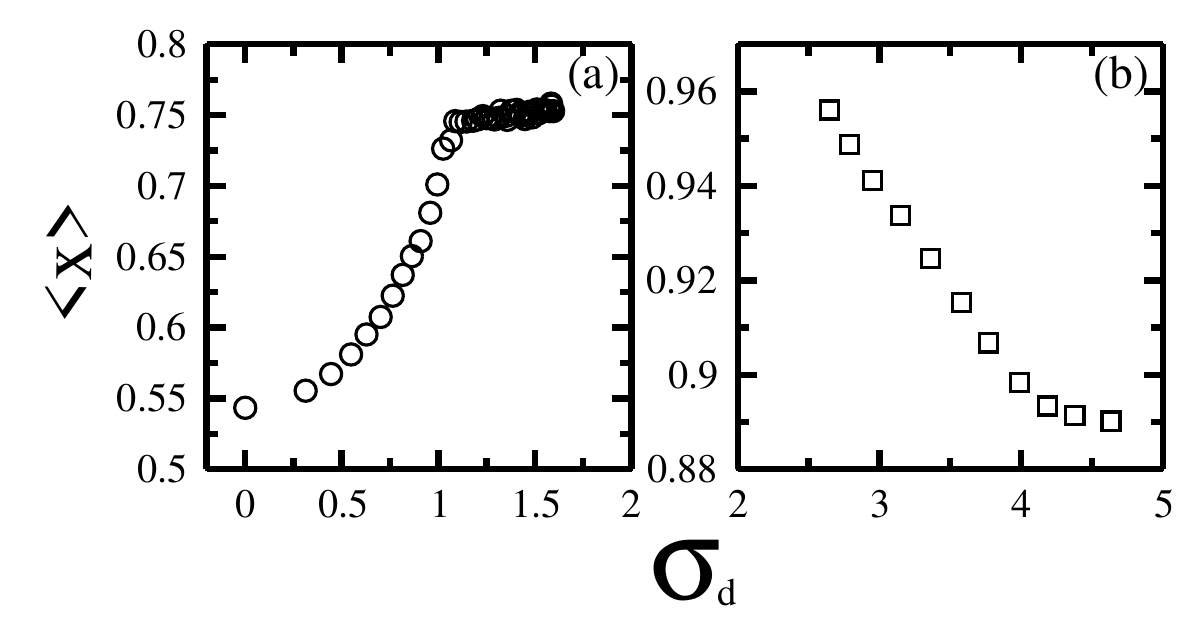}
\caption{\label{fig:epsart}The averaged investment in the final steady state as a function of $\sigma_{d}$ in the (a) FDTS network for $\psi=0.001$ (circles) and (b) dynamic network for $\psi=0.5$ (squares), averaged over 10 runs. $B(x)=-1.4x^2+6x$, $C(x)=-1.6x^2+4.56x$.}
\end{figure}

To quantify the relationship between the average investment and the degree distribution, in Fig.8 (a) and (b) we give the averaged investment in the final steady state as a function of the standard deviation $\sigma_{d}$ of the degree distribution on the FDTS network and the dynamic network respectively. It is observed that, on the FDTS network, $<x>$ firstly increases with the rise of $\sigma_{d}$. As the average investment reaches $<x>\sim0.75$, it no longer changes with the rise of $\sigma_{d}$. On the dynamic network, $<x>$ decreases monotonically with the rise of $\sigma_{d}$. A linear relationship $<x>=a\sigma_{d}+b$, where $a\sim-0.043$ and $b\sim1.068$, is found.

The averaged investment represents the level of cooperation in the mixed population and the standard deviation $\sigma_{d}$ of the degree distribution represents the homogeneity or inhomogeneity of the individual connections. The simulation results indicate that the broadening of connectivity distribution is not always beneficial for cooperation. On a dynamic network, the Poisson-like network structure is more beneficial for cooperators than the exponential-like network structure.

\section{Theoretical analysis}\label{sec:analysis}
\subsection{\label{subsec:levelA}Improved cooperation on static networks with inhomogeneous connectivity}

Firstly, let us consider how the broadening of the degree distribution affects the evolutionary branching process.

Just as that in the well-mixed case in ref.\cite{doebeli}, the evolution of investment x on the RRG can be analyzed using the mean field theory. In the original CSG, the equilibrium frequency $\rho$ of x is determined by the equation $\rho$P(x,x)+(1-$\rho$)P(x,y)=$\rho$P(y,x)+(1-$\rho$)P(y,y). On the RRG network, all the agents have the same degree $<k>$. After the investments have converged into the singular point $x^*$, evolutionary branching will occur on condition that the coexisting strategies $x_{i}$ and $x_{j}$ ($x_{i}<x^*<x_{j}$) satisfy the equation

\begin{equation}
(k_{i}-n)P(x_{i},x_{i})+nP(x_{i},x_{j})=nP(x_{j},x_{j})+(k_{j}-n)P(x_{j},x_{i}),
\end{equation}
where $k_{i}$ (or $k_{j}$) is the degree of agent i (or j) and $k_{i}=k_{j}$ and $n$ (or $k_{i}-n$) is the number of agents connected to agent i with investment $x_{j}$ (or $x_{i}$). Suppose $k_{i}=2n$. Because $P(x_{i},x_{j})=B(x_{i}+x_{j})-C(x_{i})$ and $P(x_{j},x_{i})=B(x_{j}+x_{i})-C(x_{j})$, equation (1) becomes

\begin{equation}
P(x_{j},x_{j})-P(x_{i},x_{i})=C(x_{j})-C(x_{i}).
\end{equation}

On the network with inhomogeneous connectivity, such as those with Poisson-like or exponential-like degree distribution, because the individual interactions are related to different local connectivity, the above mean-field analysis becomes difficult. To simplify the theoretical analysis, in the following, we suppose only two kinds of investment $x_{i}$ with degree $k_{i}$ and $x_{j}$ with degree $k_{j}$ exist and they are evenly distributed in the population. The wealth of agent i and agent j can be written as

\begin{equation}
w_{i}=T[(k_{i}-n')P(x_{i},x_{i})+n'P(x_{i},x_{j})],
\end{equation}

\begin{equation}
w_{j}=T[(k_{j}-n'')P(x_{j},x_{j})+n''P(x_{j},x_{i})],
\end{equation}
where $n'$(or $n''$) is the number of agents connected to agent i (or agent j) with investment $x_{j}$ (or $x_{i}$). Suppose $k_{i}=2n-k_{\alpha}$, $k_{j}=2n+k_{\alpha}$, and $x_{i}$ and $x_{j}$ are well-distributed as an agent's immediate neighbors, that is, $n'=\frac{k_{i}}{2}$, $n''=\frac{k_{j}}{2}$, equations (3) and (4) become

\begin{equation}
w_{i}=T(n-\frac{k_{\alpha}}{2})[P(x_{i},x_{i})+P(x_{i},x_{j})],
\end{equation}

\begin{equation}
w_{j}=T(n+\frac{k_{\alpha}}{2})[P(x_{j},x_{j})+P(x_{j},x_{i})].
\end{equation}
Combine equations (2), (5) and (6), we obtain

\begin{equation}
w_{j}-w_{i}=\frac{Tk_{\alpha}}{2}[P(x_{j},x_{j})+P(x_{j},x_{i})+P(x_{i},x_{i})+P(x_{i},x_{j})]>0,
\end{equation}
which indicates that the equilibrium of the coexistence of $x_{i}$ and $x_{j}$ on the RRG is destroyed on the network with inhomogeneous connectivity. The larger the value of $k_{\alpha}$, the larger the difference in
wealth.

Then, let's see how the adaptive dynamics continue evolving after the above state has been reached. Equation (7) implies that, on the network with inhomogeneous connectivity, no matter what the investment is, the agent with a higher degree keeps its state while the agent with a lower degree learns from it. So that the clusters of the agents with the same investment emerge. Suppose agent $i$ in cluster $x_{i}$ meets agent $j$ in cluster $x_{j}$, the payoffs are

\begin{equation}
P_{i}=(k_{i}-1)[B(2x_{i})-C(x_{i})]+B(x_{i}+x_{j})-C(x_{i}),
\end{equation}

\begin{equation}
P_{j}=(k_{j}-1)[B(2x_{j})-C(x_{j})]+B(x_{i}+x_{j})-C(x_{j}).
\end{equation}
On the condition $k_{i}=k_{j}$, $x_{i}<x_{j}$ and $P(x_{j},x_{j})-P(x_{i},x_{i})>C(x_{j})-C(x_{i}$), we obtain

\begin{equation}
P_{j}-P_{i}=(k_{i}-1)\{[B(2x_{j})-C(x_{j})]-[B(2x_{i})-C(x_{i})]\}+[C(x_{i})-C(x_{j})]>0.
\end{equation}
Equation (10) implies that the agents in cluster $x_{i}$ will learn from the agents in cluster $x_{j}$, so that the cluster with a higher investment is stable while the cluster with a lower investment is unstable. Therefore, on the network with inhomogeneous connectivity, there will be more agents adopting a higher investment in the final steady state.

\subsection{\label{subsec:levelB}Improved cooperation on a dynamic network}

On the static network, although the broadening of the degree distribution improves cooperation, the average investment can not reach the highest level because of the existence of the agents with a higher degree and a lower investment\cite{zhong4}. On the dynamic network, such a limitation is relaxed and the cooperation is further improved.

Firstly, consider the evolution of cluster $x_{i}$ and $x_{j}$ ($x_{i}<x_{j}$) on the dynamic network. Let $w_{im}$ and $w_{jn}$ be the wealth of agent $m$ and agent $n$ connected to agent $i$ and agent $j$ and having lower degrees $k_{m}$ and $k_{n}$ respectively.

\begin{equation}
w_{im}=Tk_{m}[B(2x_{i})-C(x_{i})],
\end{equation}

\begin{equation}
w_{jn}=Tk_{n}[B(2x_{j})-C(x_{j})].
\end{equation}
For $k_{m}=k_{n}$ and $B(2x_{i})-C(x_{i})<B(2x_{j})-C(x_{j})$, we obtain

\begin{equation}
w_{im}<w_{jn}.
\end{equation}

For $w_{jn}=<w>$, $w_{im}$ is less than $<w>$. Therefore, compared with agent $n$, agent $m$ is more possible to rewire its link(s) and the degree of agent $i$ will decrease accordingly. The decrease in degree leads to the drop of the wealth of agent $i$ and it may learn from other agents with a higher wealth. The agent with a higher degree and a higher investment is more possible to be learned from and the average investment of the population will increase accordingly. Therefore, the level of cooperation on a dynamic network is higher than that on a static network.

Then, consider why the level of cooperation on the dynamic network with a Poisson-like degree distribution is higher than that on the dynamic network with an exponential-like degree distribution. For the network with degree distribution $P(k)=\delta(k-<k>)$, the wealth of each agent is the same as $<w>$. For a Poisson degree distribution, the ratio of the agents whose wealth is lower than $<w>$ is $\frac{1}{2}$ whereas it is $\frac{\int_{1}^{<k>}e^{-k}dk}{\int_{1}^{\infty}e^{-k}dk}=1-\frac{1}{e^{<k>-1}}$ for an exponential degree distribution. For $<k>\ge2$, we get $1-\frac{1}{e^{<k>-1}}>\frac{1}{2}$. Therefore, there are more agents who are constantly rewiring their links on the network with an exponential-like degree distribution than that on the network with a Poisson-like degree distribution.

Finally, let's have a look at how the average investment changes with the ratio of the agents constantly rewiring their links. We divide the population into three groups: group A with investment $x_{a}$, group B with investment $x_{b}$ and group C with investment $x_{c}$. Suppose $x_{a}>x_{b}>x_{c}$ and the leaders in the three groups are agent A, agent B and agent C respectively. Originally, the number of the agents with a lower degree connected to A, B, C is equal, $n_{A}=n_{B}=n_{C}=\frac{N}{3}$. If only $\frac{N}{3}$ of the agents rewire their links, they should be the agents with the lowest investment $x_{c}$ because of their lowest wealth. They may reconnect to A, B, C equally and $n_{A}:n_{B}:n_{C}=4:4:1$. Compared with that in the original case, we find the rise of the average investment. If there are $\frac{2N}{3}$ of the agents rewire their links, they should be the agents with investment $x_{c}$ and $x_{b}$ and $n_{A}:n_{B}:n_{C}$ becomes $5:2:2$. If all the agents rewire their links, $n_{A}:n_{B}:n_{C}=1:1:1$ and the average investment is the same as that in the first case. The above analysis shows that, as we increase the ratio of the agents constantly rewiring their links, the average investment firstly increases and then decreases. Therefore, a larger fluctuation of cooperator clusters means a lower average investment, which may be the reason for the decrease of the average investment on the dynamic network with an exponential-like degree distribution.

Compared with that on the network with a Poisson-like degree distribution, there are more agents who are constantly rewiring their links on the network with an exponential-like degree distribution. In the present model, a higher level of cooperation is easier to be attained on the dynamic network with a Poisson-like degree distribution than that with an exponential-like degree distribution.

\section{Summary}\label{sec:summary}

We study the level of cooperation of the CSG on a dynamic network. Compared with the case on a static network, rewiring leads to the rise of cooperation in the present model. The role of degree distribution in the improvement of cooperation is extensively studied. Contrary to the findings on the static network, an exponential-like degree distribution is found to suppress cooperation in the coevolutionary dynamics. The theoretical analysis indicates that the positively biased pay-off structure brings about the unequal allocation of the wealth among the agents with different degrees. A large difference in degree leads to a large difference in wealth and thereafter the extreme instability of cooperator clusters, which results in the suppression of cooperation on the network with an exponential-like degree distribution in contrast to the network with a Poisson-like degree distribution.

The present work shows that, even if without an additional mechanism, such as the increased cost, in the coevolutionary process, because of the existence of the agents with a higher degree and a low investment, the advantage of the broadening of the degree distribution in the improvement of cooperation may be lost and the agent with a higher degree may become the inhibitor of cooperation . In the present model, we have only discussed the replicator dynamics of investment x but not the individual relinking preferences. In reality, it is possible that agent i becomes agent j's identical offspring, including both its investment and relinking preference. In the future, the coupled dynamics will be studied extensively in such a case and the difference between the roles of dynamic and static networks in the change of system behavior is the main interest of us.

\section*{Acknowledgments}
This work is the research fruits of the Humanities and Social Sciences Fund sponsored by Ministry of Education of China (Grant No. 10YJAZH137), Natural Science Foundation of Zhejiang Province (Grant No. Y6110687), Social Science Foundation of Zhejiang Province (Grant No. 10CGGL14YB) and National Natural Science Foundation of China (Grant No. 10805025).





\bibliographystyle{model1-num-names}

\begin{thebibliography}{00}

\bibitem{nowak}M.A. Nowak, Science 314 (2006) 1560.

\bibitem{wang}W.X. Wang, J. Ren, G.R. Chen, B.H. Wang, Phys. Rev. E 74 (2006) 056113.

\bibitem{zhong1}L.X. Zhong, D.F. Zheng, B. Zheng, P.M. Hui, Phys. Rev. E 72 (2005) 026134.

\bibitem{zheng}D.F. Zheng, H.P. Yin, C.H. Chan, P.M. Hui, Europhys. Lett. 80 (2007) 18002.

\bibitem{fu}F. Fu, C. Hauert, M.A. Nowak, L. Wang, Phys. Rev. E 78 (2008) 026117.

\bibitem{hauert}C. Hauert, A. Traulsen, H. Brandt, M.A. Nowak, K. Sigmund, Science 316 (2007) 1905.

\bibitem{watts}J. Watts, S.H. Strogatz, Nature 393 (1998) 440.

\bibitem{albert}R. Albert, A.-L. Barabasi, Rev. Mod. Phys. 74 (2002) 47.

\bibitem{ivanov}P.Ch. Ivanov, D.Y. Ma Qianli, R.P. Bartsch, J.M. Hausdorff, Lu¨ªs A. Nunes Amaral, V. Schulte-Frohlinde, H.E. Stanley, M. Yoneyama, Phys. Rev. E 79 (2009) 041920.

\bibitem{masucci}A.P. Masucci, G.J.Rodgers, Physica A 387 (2008) 3781.

\bibitem{baek}S.K. Baek, B.J. Kim, Phys. Rev. E 78 (2008) 011125.

\bibitem{ohtsuki}H. Ohtsuki, C. Hauert, E. Lieberman, M.A. Nowak, Nature 441 (2006) 502.

\bibitem{szolnoki}A. Szolnoki, G. Szabo, Europhys. Lett. 77 (2007) 30004.

\bibitem{zhong2}L.X. Zhong, D.F. Zheng, B. Zheng, C. Xu, P.M. Hui, Europhys. Lett. 76 (2006) 724.

\bibitem{chan}C.H. Chan, H.P. Yin, P.M. Hui, D.F. Zheng, Physica A 387 (2008) 2919.

\bibitem{masuda}N. Masuda, Proc. R. Soc. B 274 (2007) 1815.

\bibitem{tanimoto}J. Tanimoto, A. Yamauchi, Physica A 389 (2010) 2284.

\bibitem{perc1}M. Perc, A. Szolnoki, Phys. Rev. E 77 (2008) 011904.

\bibitem{perc2}M. Perc, New J. Phys. 8 (2006) 22.

\bibitem{johnson}N.F. Johnson, C. Xu, Z. Zhao, N. Ducheneaut, N. Yee, G. Tita, and P.M. Hui, Phys. Rev. E 79 (2009) 066117.

\bibitem{xu}C. Xu, P. M. Hui, Y. Y. Yu, and G. Q. Gu, Physica A 388 (2009) 4445.

\bibitem{zhong3}L.X. Zhong, T. Qiu, F. Ren, P.P. Li, and B.H. Chen, Europhys. Lett. 94 (2011) 18004.

\bibitem{eguiluz}Victor M. Eguiluz, Claudio J. Tessone, Advances in Complex Systems 12 (2009) 221.

\bibitem{sinatra}R. Sinatra, J. Iranzo, Jesus Gomez-Gardenes, Luis M. Floria, V. Latora, Y. Moreno, J. Stat. Mech. (2009) 09012.

\bibitem{li}X. Li, L. Cao, Phys. Rev. E 80 (2009) 066101.

\bibitem{iranzo}J. Iranzo, J. Roman, A. Sanchez, J. Theor. Biol. 278 (2011) 1.

\bibitem{szolnoki1}A. Szolnoki, M. Perc, Europhys. Lett. 86 (2009) 30007.

\bibitem{szolnoki2}A. Szolnoki, M. Perc, Z. Danku, Europhys. Lett. 84 (2008) 50007.

\bibitem{szolnoki3}A. Szolnoki, M. Perc, New J. Phys.11 (2009) 093033.

\bibitem{perc3}M. Perc, A. Szolnoki, BioSystems 99 (2010) 109.

\bibitem{doebeli}M. Doebeli, C. Hauert, T. Killingback, Science 360 (2004) 859.

\bibitem{zhong4}L.X. Zhong, T. Qiu, J.R. Xu, Chin.Phys.Lett. 25 (2008) 2315.

\bibitem{zhong5}L.X. Zhong, F. Ren, T. Qiu, J.R. Xu, B.H. Chen, C.F. Li, Physica A 389 (2010) 2557.

\bibitem{zschaler}G. Zschaler, A. Traulsen, T. Gross, New J. Phys. 12 (2010) 093015.

\bibitem{chen}X. Chen, F. Fu, L. Wang, Phys. Rev. E 80 (2009) 051104.

\bibitem{assenza}S. Assenza, J.G. Gardefles, V. Latora, Phys. Rev. E 78 (2008) 1431.





\end{thebibliography}



\end{document}